\documentclass[journal=apchd5,manuscript=article]{achemso}

\usepackage[version=3]{mhchem} 
\usepackage{graphicx}
\usepackage{amsmath,amssymb}
\usepackage{upgreek}
\usepackage{units}
\usepackage{booktabs}
\usepackage{xcolor}

\author{Shreyas Ramachandran}
\author{Sim\~ao Jo\~ao}
\author{Hanwen Jin}
\author{Johannes Lischner}
\email{j.lischner@imperial.ac.uk}
\affiliation[Imperial College London] {Department of Materials, Imperial College London, South Kensington Campus, London SW7 2AZ, United Kingdom}
\alsoaffiliation[Thomas Young Centre]{The Thomas Young Centre for Theory and Simulation of Materials, London E1 4NS, United Kingdom}

\title[]{Hot Carriers from Intra- and Interband Transitions in Gold-Silver Alloy Nanoparticles}

\abbreviations{IR,NMR,UV}

\begin{document}


\begin{abstract}
Hot electrons and holes generated from the decay of localized surface plasmons in metallic nanoparticles can be harnessed for applications in solar energy conversion and sensing. In this paper, we study the generation of hot carriers in large spherical gold-silver alloy nanoparticles using a recently developed atomistic modelling approach that combines a solution of Maxwell's equations with large-scale tight-binding simulations. We find that hot-carrier properties depend sensitively on the alloy composition. Specifically, nanoparticles with a large gold fraction produce hot carriers under visible light illumination while nanoparticles with a large silver fraction require higher photon energies to produce hot carriers. Moreover, most hot carriers in nanoparticles with a large gold fraction originate from interband transitions which give rise to energetic holes and "cold" electrons near the Fermi level. Increasing the silver fraction enhances the generation rate of hot carriers from intraband transitions which produce energetic electrons and "cold" holes. These findings demonstrate that alloy composition is a powerful tuning parameter for the design of nanoparticles for applications in solar energy conversion and sensing that require precise control of hot-carrier properties.
\end{abstract}

\section{Introduction}
Plasmonic nanoparticles (NPs) are an emerging platform for solar energy conversion devices, such as solar cells or photocatalysts. These systems feature large light absorption cross sections due to localized surface plasmons (LSPs) which are collective oscillations of the conduction electrons in the NP~\cite{maier_plasmonics_2007}. On the time scale of tens of femtoseconds, the LSP decays with the dominant decay channel being the Landau damping mechanism which results in the excitation of electron-hole pairs~\cite{brown_nonradiative_2016}. The excited carriers have large energies compared to the available thermal energy ~\cite{narang_plasmonic_2016}.  
Such energetic or "hot" carriers can be harnessed for a wide range of applications: for example, they can be used to drive chemical reactions~\cite{yan_quantum_2016, mukherjee_hot_2013}, transferred to semiconductors for photovoltaic applications~\cite{takahashi_solid_2011, enrichi_plasmonic_2018}, or used for photodetection~\cite{tang_plasmonic_2020, sun_photodetection_2019}.

Nanoparticles composed of noble metals, such as Au~\cite{huynh_effect_2022, salmon-gamboa_rational_2021, pensa_spectral_2019, berdakin_interplay_2020} and Ag~\cite{hoffmann_conjugated_2022, huynh_effect_2022, stefancu_fermi_2021, seemala_plasmon-mediated_2019, berdakin_interplay_2020}, have been studied intensively because of their favourable optical properties. However, pure metal NPs have some disadvantages: in particular, the LSP energy of Ag nanoparticles is quite high resulting in a small overlap with the solar spectrum. In contrast, Au NPs have a lower LSP energy, but a large fraction of the photoexcited carriers arises from interband transitions which do not generate energetic electrons that can be used for reduction reactions, such as CO$_2$ reduction. To overcome the limitations of the pure materials, several studies investigated the properties of noble metal alloys~\cite{cortie_synthesis_2011, link_alloy_1999, blaber_review_2010,jawad_plasmonic_2019,darmadi_optimization_2021,bhatia_tunable_2020,coviello_recent_2022,gong_noble_2016}. Because of the disorder in these systems, the rate of intraband transitions which give rise to energetic electrons is enhanced. This was exploited by Valenti and coworkers who demonstrated that Ag-Au alloy nanoparticles exhibit higher hot-electron injection efficiencies into the TiO$_2$ substrate compared to pure Au and Au nanoparticles~\cite{valenti_hot_2017}. Dahiya and coworkers prepared various Au-Cu alloys with tunable optical properties and higher catalytic turnover frequency than pure Au~\cite{dahiya_plasmonic_2022}.

To accelerate the design of plasmonic alloy NPs for application in solar energy conversion devices, a detailed understanding of their electronic structure is required. However, application of standard electronic structure techniques, such as ab initio density-functional theory (DFT), to plasmonic NPs is challenging because of their large size containing thousands or even millions of atoms~\cite{fojt_tailoring_2023, rossi_hot-carrier_2020, fojt_hot-carrier_2022, kumar_plasmon-induced_2019, brown_nonradiative_2016}. As a consequence, simplified electronic structure techniques, such as jellium or spherical well models, are often used to describe large NPs~\cite{manjavacas_plasmon-induced_2014, roman_castellanos_dielectric_2021, roman_castellanos_generation_2020}, but these techniques do not capture d-band derived NP states~\cite{tagliabue_ultrafast_2020} or the dependence of electronic properties on the exposed NP facets~\cite{rossi_hot-carrier_2020}. To overcome these problems, Jin and coworkers recently introduced an atomistic electronic structure technique based on the tight-binding approach that allows the modelling of hot-carrier generation in large NPs containing more than one million atoms~\cite{jin_plasmon-induced_2022}. They used this new approach to study the contributions to the hot-carrier generation rate from interband and intraband transitions in spherical NPs of Au, Ag and Cu~\cite{jin_plasmon-induced_2022} and also investigated hot-carrier generation in bi-metallic photocatalysts of Au and Pd~\cite{jin_theory_2023}.

In this paper, we study hot-carrier generation in Ag-Au alloy NPs using the atomistic modelling approach developed by Jin and coworkers~\cite{jin_plasmon-induced_2022}. We find that hot-carrier generation rates depend sensitively on the material composition and also on the photon energy. Specifically, a decrease in the Au fraction $x_\text{Au}$ increases the LSP energy and also enhances the excitation rate of hot carriers generated from intraband transitions relative to the rate generated from interband transitions. Our results demonstrate that the alloy composition is an important tuning parameter for designing plasmonic NPs for solar energy conversion applications and our approach can be applied to other alloy systems in the future.

\section{Results and discussion}

\begin{figure}[H]
    \centering
    \includegraphics[width=0.35\linewidth]{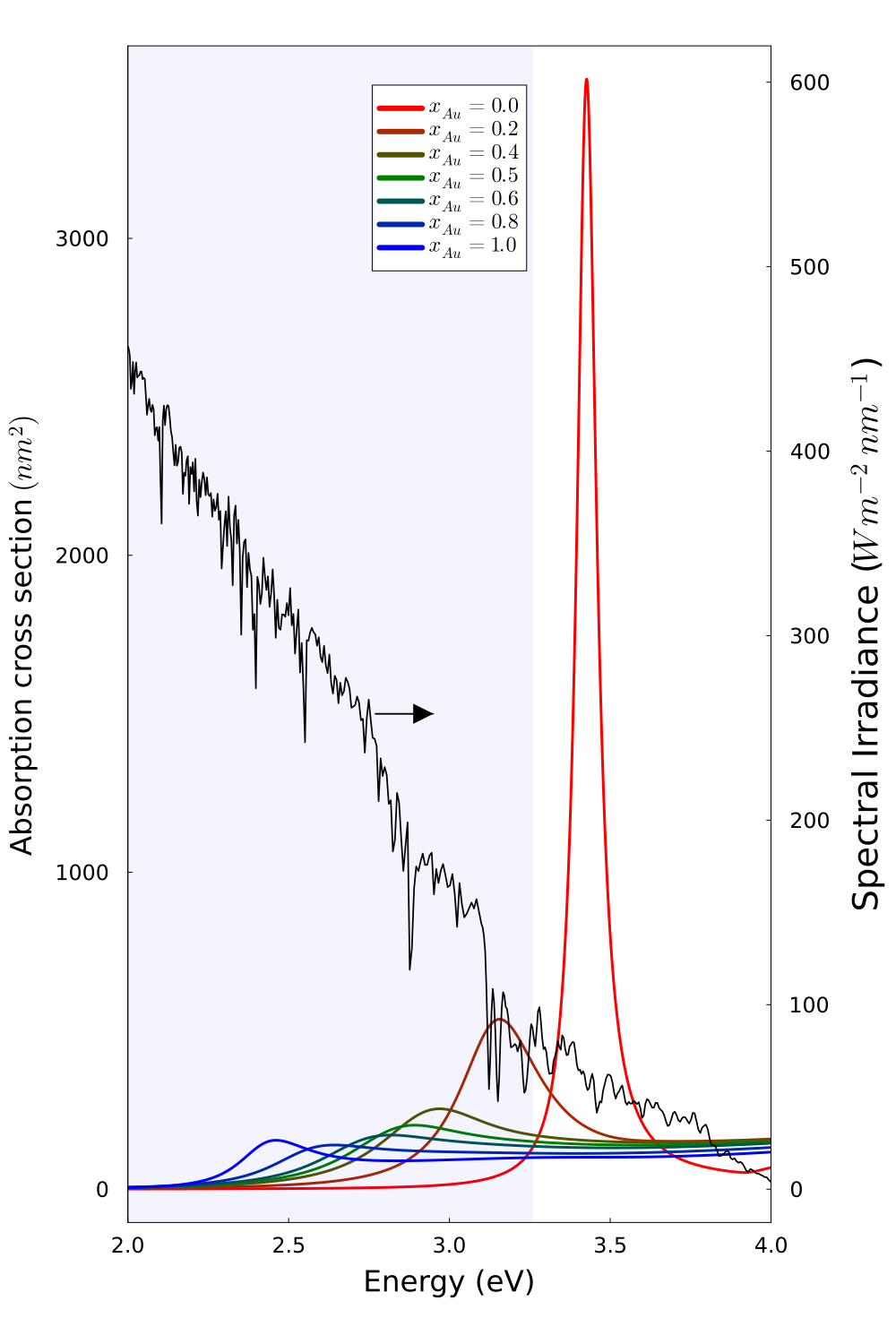}
    \caption{Absorption cross sections of spherical Au-Ag alloy nanoparticles of 20 nm diameter as a function of the NP composition. The solar irradiance (AM1.5 spectrum) as a function of photon energy is shown in black with the visible region shaded in blue. The cross sections were calculated using the quasistatic approximation with experimentally measured bulk dielectric functions, see Methods section.}
    \label{fig:cross}
\end{figure}

Figure~\ref{fig:cross} shows the absorption cross sections of spherical Au-Au alloy nanoparticles of 20 nm diameter for a range of compositions $x_{\text{Au}}=(0.0, 0.2, 0.4, 0.6, 0.8, 1.0)$. For pure Ag nanoparticles (corresponding to $x_{\text{Au}}=0.0$), the cross section exhibits a sharp peak at the LSP energy of $\sim 3.4$~eV. Note that this peak lies outside the visible region and has a small overlap with the solar spectrum. Therefore, pure Ag NPs are not attractive for solar energy conversion devices, such as photocatalysts or solar cells. As the Ag content is reduced, the LSP energy decreases monotonically until the LSP energy of pure gold ($\sim 2.4$~eV) is reached. As the Au content increases, the LSP moves into the visible region and the absorption cross section has a larger overlap with the solar spectrum. Strikingly, the LSP peak of the alloy NPs is significantly smaller and broader than the Ag NP peak. These findings demonstrate the tunability of the optical properties of NPs by alloying. 

\begin{figure}[H]
    \centering
    \includegraphics[width=0.95\linewidth]{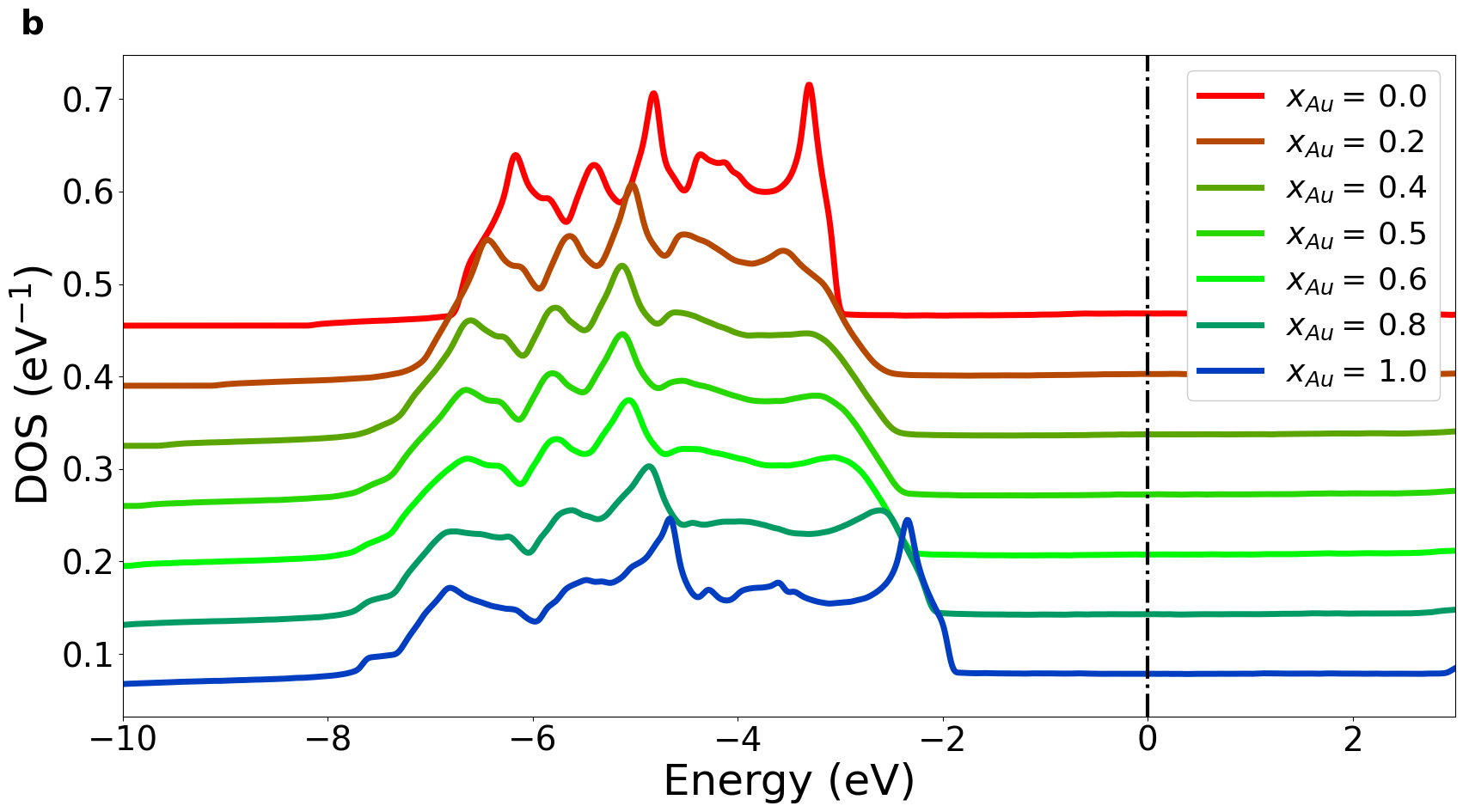}
    \caption{Normalized DOS for different spherical Au-Ag alloy nanoparticles of 20 nm diameter. The DOS is normalized to the total number of states available in the alloy NP ($9\times N_{atoms}$). The different curves are offset by 0.065 eV for clarity and the Fermi level is set to 0 eV.}
    \label{fig:dos}
\end{figure}

Figure~\ref{fig:dos} shows the density of states (DOS) of spherical Ag-Au alloy nanoparticles (20 nm diameter corresponding to 246,857 atoms) for a range of compositions. Below the Fermi energy, the DOS exhibits a prominent feature arising from d-band states. The onset of this d-band peak is at $-1.9$~eV for pure Au and at $-3.1$~eV for pure Ag. For the pure materials, the onset of the DOS is characterized by a sharp peak. In contrast, the alloy NPs do not exhibit such a sharp peak at the d-band onset. Such a broadening has also been observed in photoemission experiments of alloy nanoparticles~\cite{valenti_hot_2017}. The broadening is a consequence of the disorder that gives rise to scattering between the unperturbed states of the pure materials. This leads to a finite lifetime of the states which is reflected by an increase in their linewidths giving rise to a broader DOS.

\begin{figure}[H]
    \centering
    \includegraphics[width=1.0\linewidth]{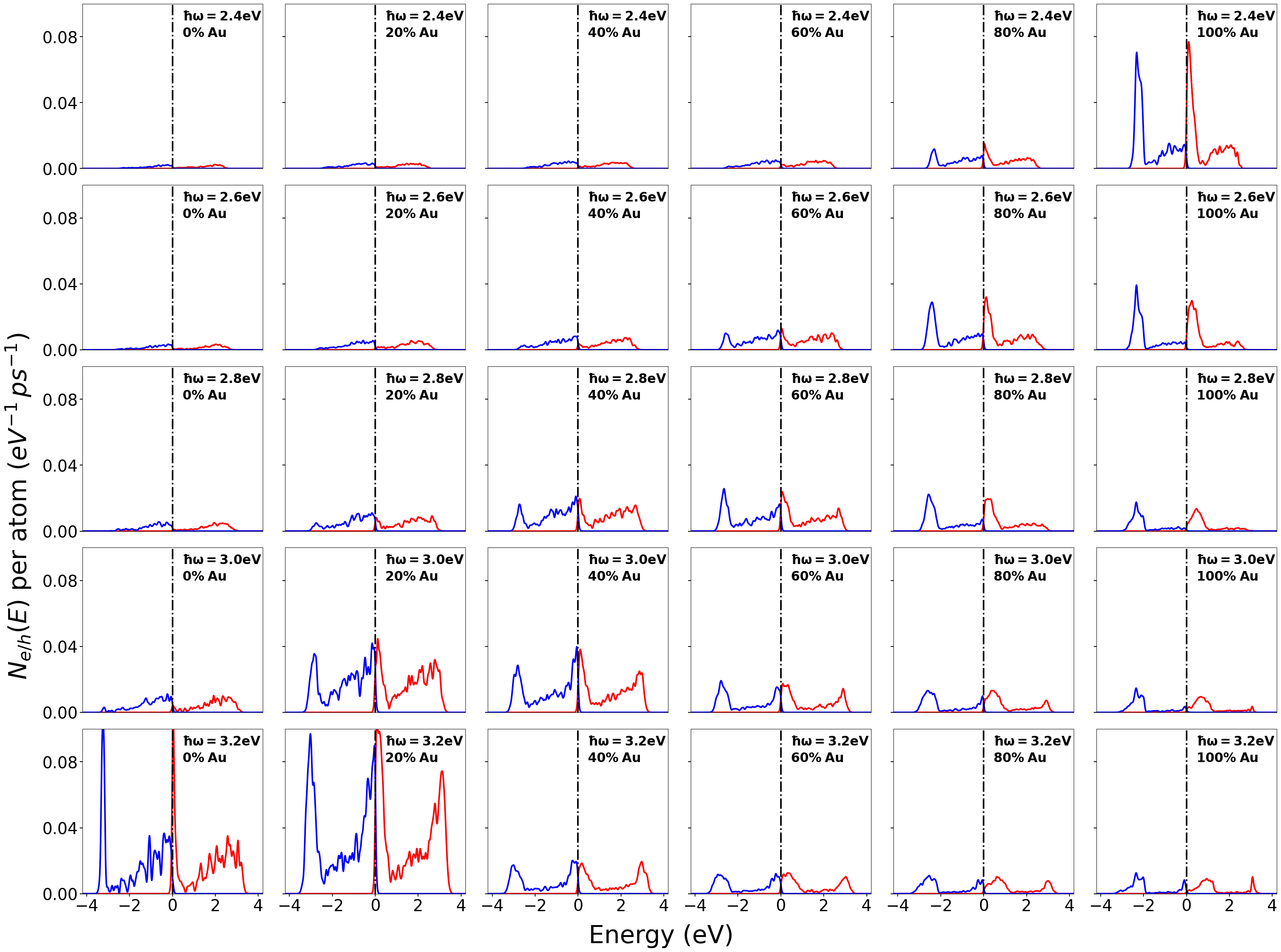}
    \caption{Hot-carrier generation rates for spherical Ag-Au alloy nanoparticles of 8 nm diameter as function of alloy composition and photon energy. Hot-electron (hot-hole) generation rates are denoted by red (blue) lines. Each column corresponds to a fixed composition and each row to a fixed photon energy. The Fermi level is set to 0 eV.}
    \label{fig:hcg_all}
\end{figure}

Figure~\ref{fig:hcg_all} shows the hot-carrier generation rates of the Ag-Au alloy nanoparticles for different photon energies ranging from 2.4~eV to 3.2~eV. Pure Au has a plasmon energy of 2.4 eV. Since this photon energy is larger than the d-band onset (relative to the Fermi level), the hot-carrier generation rate exhibits sharp peaks arising from interband transitions. Interband transitions give rise to energetic holes (peak near $-2$~eV), but the electrons are "cold" (peak near the Fermi level). In addition, a smaller and broader peak arising from intraband transitions can be observed. Such transitions give rise to energetic electrons (peak near $+2$~eV), but cold holes (peak near the Fermi level). Increasing the photon energy leads to a reduction of the hot-carrier generation rates in pure Au since away from the plasmon resonance the field enhancement inside the NP is weaker. 

For an alloy NP with $x_{\text{Au}}=0.8$, the largest hot-carrier generation rate is found at a photon energy of 2.6 eV corresponding to the LSP energy at this composition, see Fig.~\ref{fig:cross}. In comparison to the pure Au result, the hot-carrier generation rate of the alloy NP at the LSP energy is significantly reduced. This is a consequence of the smaller absorption cross section of alloy NPs. Interestingly, the alloy hot-carrier generation rate at 2.6 eV has a larger intraband peak compared to the pure Au result at the same photon energy. Intraband transitions in a pure Au NP are only possible because the NP surface breaks translational invariance. In an alloy, disorder breaks translational invariance already in the bulk material and therefore no surface is required to excite intraband transitions. 
This demonstrates that the relative importance of interband and intraband transitions for hot-carrier generation rates can be effectively controlled by tuning the alloy composition. 

Further reduction of the Au content to $x_{\text{Au}}=0.6$ completely suppresses the interband peak in the hot-carrier generation rates at small photon energies. This is a consequence of the shift and broadening of the d-band onset which make it increasingly difficult to excite interband transitions. For example, for NPs with $x_{Au}=0.2$, the interband peaks are only observed for photon energies exceeding 2.8 eV. 

Finally, pure Ag nanoparticles exhibit interband peaks only at photon energies larger than 3.1 eV. At photon energies closer to the LSP energy (3.4 eV for pure Ag), the hot-carrier generation rate is larger than that of pure Au nanoparticles. This is a consequence of the large absorption cross section of Ag nanoparticles, see Fig.~\ref{fig:cross}. 

\begin{figure}[H]
    \centering
    \includegraphics[height=.3\textheight]{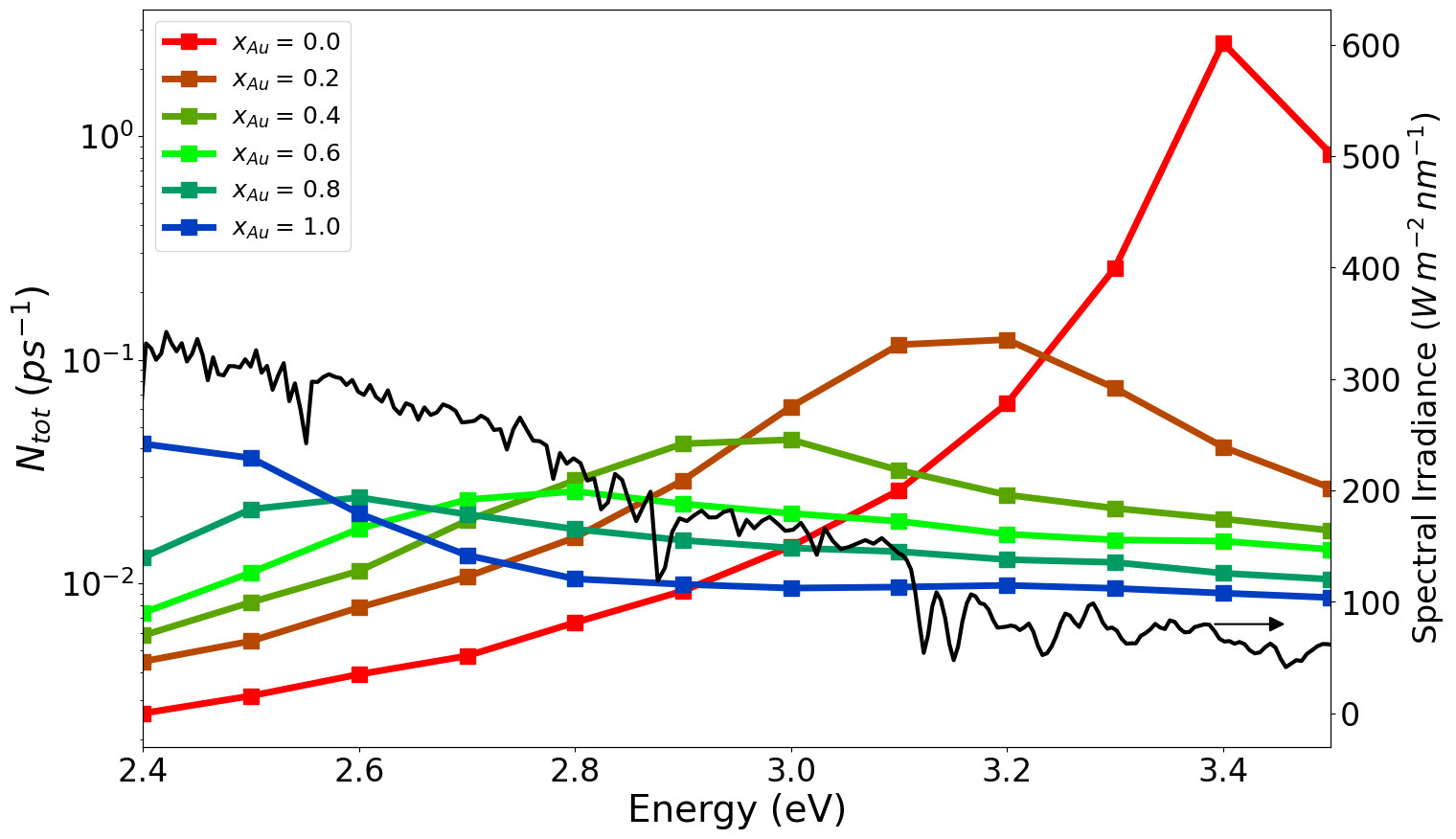}
    \caption{Total hot-carrier generation rate per atom $N_{tot}(\omega)$ for spherical Ag-Au alloy nanoparticles as a function of photon energy and alloy composition $x_\text{Au}$. The solar irradiance (AM1.5 spectrum) as a function of photon energy is shown in black.}
    \label{fig:N_total}
\end{figure}

Figure~\ref{fig:N_total} shows the total hot-carrier generation rate $N_\text{tot}(\omega)$ as function of alloy composition and photon energy. The largest generation rate is found for pure Ag NPs at their LSP energy of 3.4 eV. However, the overlap of the hot-carrier generation rate with the solar spectrum is small. The total hot-carrier generation rate of the alloy NPs is smaller and exhibits a broader peak than the pure Ag NP reflecting the behaviour of their absorption cross section. However, alloying increases the overlap of $N_\text{tot}(\omega)$ with the solar spectrum thus making alloy NPs more attractive for solar energy conversion applications.

\begin{figure}[H]
    \centering
     \includegraphics[height=.28\textheight]{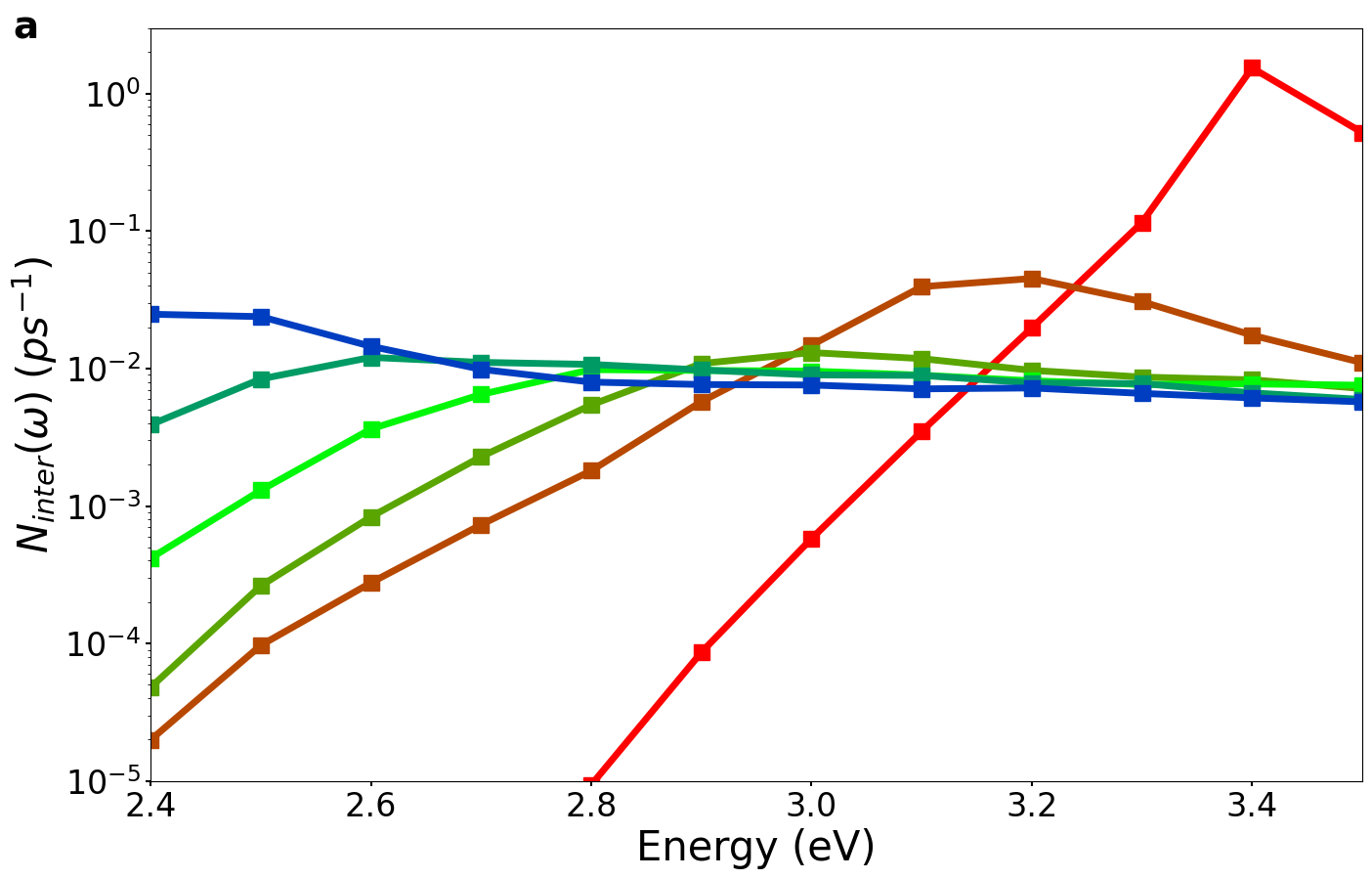}
    \includegraphics[height=.28\textheight]{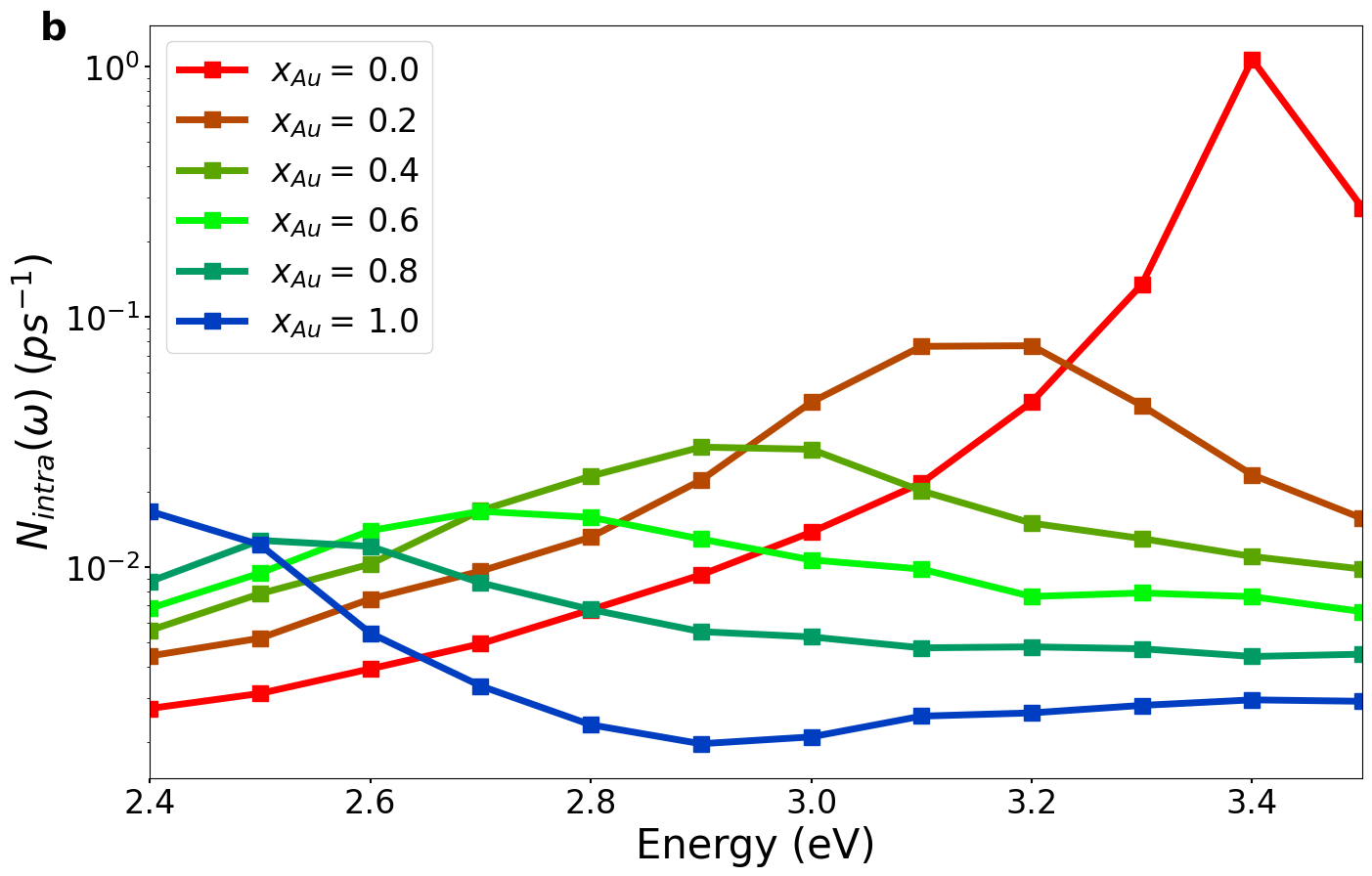}
    \includegraphics[height=.28\textheight]{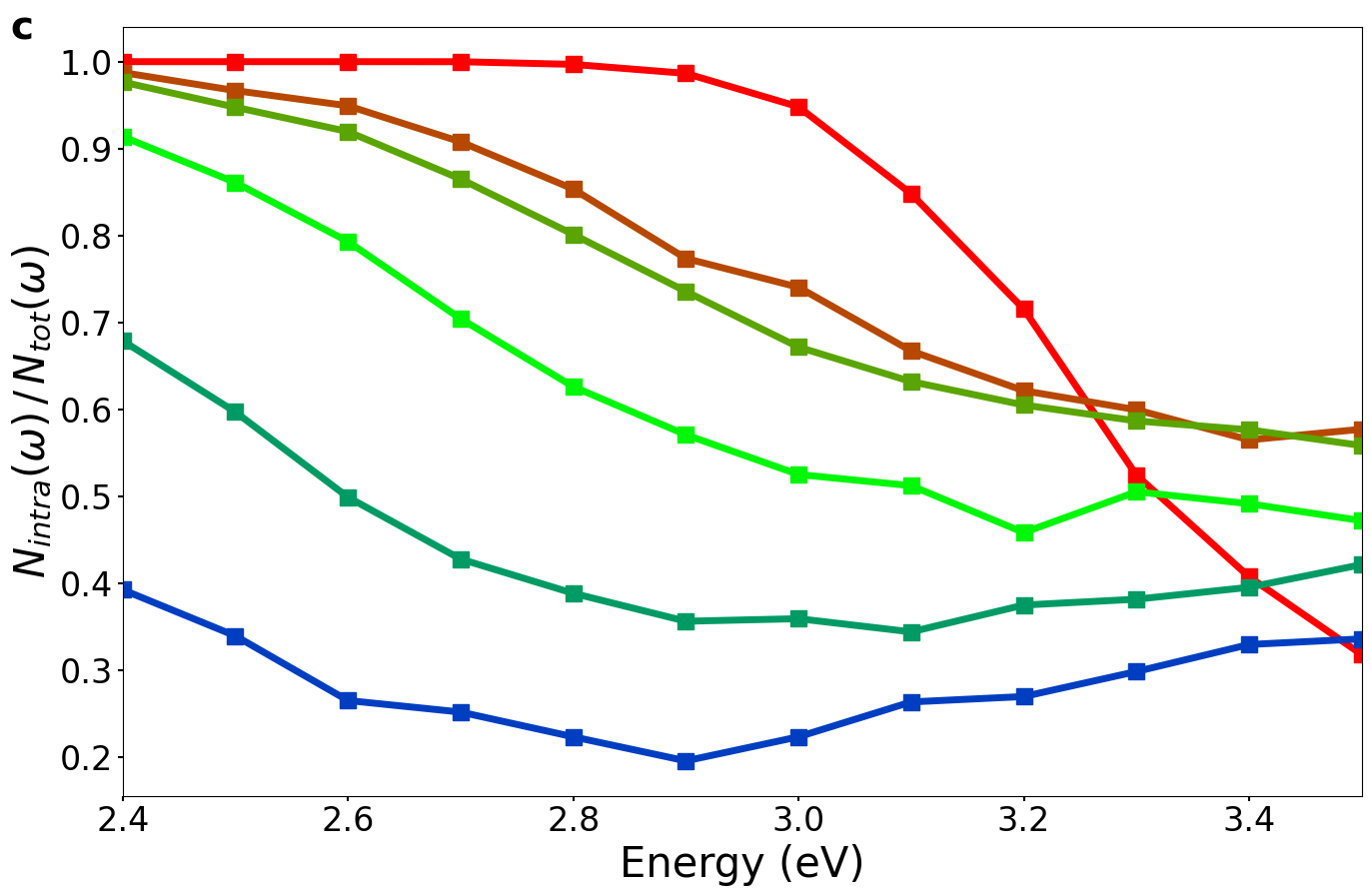}
    \caption{Contribution to the total hot-carrier generation rate per atom from interband transitions (a) and intraband transitions (b) for spherical Ag-Au nanoparticles as a function of alloy composition $x_\text{Au}$ and photon energy. (c) Ratio of intraband contribution to the total hot-carrier generation rate.}
   \label{fig:N_inter}
\end{figure}

Figure~\ref{fig:N_inter} compares the contributions from interband and intraband transitions to the total hot-carrier generation rate. Figs.~\ref{fig:N_inter}(a) and (b) show that pure Ag NPs exhibit the highest generation rate for hot carriers from both inter- and intraband transitions at the LSP energy. However, both contributions drop sharply as the photon energy is reduced. In contrast, the interband contribution in alloy NPs is relatively constant over a significant range of photon energies. Interestingly, the interband hot-carrier generation rate appears to only weakly depend on the alloy composition.  The intraband contribution peaks at the LSP energy for each NP and shrinks as the Au content increases. Finally, Fig.~\ref{fig:N_inter}(c) shows the ratio of hot carriers generated from intraband transitions to the total generation rate. It can be observed that this ratio depends sensitively both on the photon energy and the alloy composition: for pure Au NPs, the ratio is relatively small (less than 0.4) and depends weakly on the photon energy. For pure Ag NPs, the ratio increases from 0.3 to 1.0 as the photon energy is reduced from 3.5 eV to 2.4 eV. Decreasing the Au content results in a significant increase in the intraband contribution. For example, NPs with $x_\text{Au}=0.2$ and $x_\text{Au}=0.4$ exhibit a dramatic increase compared to pure Au NPs in the relative contribution of hot carriers from intraband transitions at small photon energies.

\section{Conclusion}
In this work, we studied hot-carrier properties of alloys of Ag-Au alloy nanoparticles using an atomistic modelling technique that combines a solution of Maxwell's equations with large-scale tight-binding simulations. Specifically, we have calculated hot-carrier generation rates of large spherical nanoparticles for different compositions and photon energies. Our results demonstrate the hot-carrier generations depend sensitively the alloy composition which controls both the optical and electronic properties of the nanoparticles: for example, the energy of the localized surface plasmon depends on the material composition which in turn determines at which photon energy most hot carriers are generated. Moreover, the alloy composition determines the onset of d-band states relative to the Fermi level which in turn controls the photon energy at which interband transitions from d-band states to sp-band states become possible. Interband transitions produce energetic holes but "cold" electrons near the Fermi level. To generate energetic electrons which can be harnessed for reduction reactions, the contribution of hot carriers from intraband transitions within the sp-band must be increased. This can be achieved by increasing the Ag content at fixed photon energy. Our findings demonstrate that the alloy composition is an important parameter which allows the continuous tuning of hot-carrier properties for specific applications in photocatalysis, photovoltaics or sensing. In the future, our approach can be applied to other alloy systems.

\section{Methods}

We calculate hot-carrier generation rates $N_e(E,\omega)$ (with $E$ being the energy of the electrons and $\omega$ is the angular frequency of light) of spherical Au-Ag alloy nanoparticles following the approach developed by Jin and coworkers~\cite{jin_plasmon-induced_2022}. Using Fermi's golden rule, $N_e(E,\omega)$ can be expressed as 

\begin{equation}
    N_e(E,\omega) = \frac{2}{V} \sum_{if} \Gamma_{if}(\omega) \delta(E-E_f;\sigma),
    \label{HCG_eqn}
\end{equation}

where $i$ and $f$ label initial and final states, respectively, $V$ is the nanoparticle volume and the factor of 2 accounts for spin degeneracy. We define $\delta(x;\sigma)=\frac{1}{\sqrt{2\pi\sigma^2}}\exp{(-\frac{x^2}{2\sigma^2})}$ which becomes a delta function in the limit of $\sigma\rightarrow0^+$ and $\sigma=0.06$~eV is a broadening parameter. Finally, $\Gamma_{if}$ is given by

\begin{equation}
    \Gamma_{if}(\omega)=\frac{2\pi}{\hbar} \Big|{\langle{\psi_f}| \hat{\Phi}_{\text{tot}}(\omega)|\psi_i\rangle}\Big|^2 \delta(E_f-E_i-\hbar\omega;\sigma)f(E_i)(1-f(E_f)),
    \label{gamma_eqn}
\end{equation}

where $f(E)$ denotes the Fermi-Dirac distribution evaluated at $T=298$~K, $\hat{\Phi}_{\text{tot}}(\omega)$ is the total potential inside the nanoparticle. The total potential is calculated within the quasistatic approximation, i.e. we solve the Laplace equation $\nabla \cdot (\epsilon(\mathbf{r},\omega)\nabla\Phi_{\text{tot}}(\mathbf{r},\omega))=0$ for spherical nanoparticles with the boundary condition at infinity $\Phi_{\text{tot}}(\mathbf{r},\omega) = \Phi_{\text{ext}}(\mathbf{r},\omega)= -E_0 z$. The analytical solution is given by

\begin{equation}
    \Phi_{\text{tot}}(\mathbf{r},\omega) = -eE_{0} \frac{3\epsilon_m}{2\epsilon_m+\epsilon(\omega)} z.
    \label{laplace_eqn}    
\end{equation}

Here, $-e$ is the electron charge, $E_0=1$~V/nm, $\epsilon_m=1$ is the dielectric constant of the environment and the dielectric function of the bulk alloy $\epsilon(\omega)$ is evaluated using the analytical model of Rioux and coworkers~\cite{rioux_analytic_2014}.

The hot-hole generation rate $N_h(E,\omega)$ is obtained by swapping the indices of the initial and final states in Eq.~\ref{HCG_eqn}.

The electronic states of the alloy nanoparticles are obtained using a tight-binding (TB) model based on the work of Hegde et al. \cite{hegde_environment-dependent_2014} that uses the valence orbitals of Ag ($4d$, $5s$, and $5p$) and Au ($5d$, $6s$, and $6p$) as basis set. The inter-atomic matrix elements of the Hamiltonian are obtained using the two-centre Slater-Koster formalism~\cite{slater_simplified_1954}. The intra-atomic matrix elements describe the onsite energies $E_{il}$ (where $i$ labels an atom and $l=s,p,d$) which are expressed as

\begin{equation}
    E_{il} = \varepsilon_{il} + \sum_{j \in \text{NN}_i} n_{ij}I_{ll\sigma}(R_{ij}), 
    \label{intra_onsite}
\end{equation}

where $\varepsilon_{il}$ is a constant and the second term on the right-hand side captures the variations of the onsite energies due to the chemical environment of atom $i$. Specifically, $\text{NN}_i$ denotes the set of nearest neighbors of atom $i$, $R_{ij}$ is the distance between atoms $i$ and $j$, $n_{ij}$ is the direction cosine between atoms $i$ and $j$, and $I_{ll\sigma}$ is given by~\cite{harrison_electronic_2012}

\begin{equation}
    I_{ll\sigma}(R_{ij}) = I^{(0)}_{ll\sigma} \exp \bigg[-p_{ll\sigma}\Big(\frac{R_{ij}}{R^{(0)}_{ij}}-1\Big)\bigg],
    \label{SK-intra}
\end{equation}

where $R^{(0)}_{ij}$ is the equilibrium distance between atoms $i$ and $j$ and $p_{ll\sigma}$ and $I^{(0)}_{ll\sigma}$ are fitting parameters. 

The parameters of the tight-binding Hamiltonian for the Ag-Au alloy are taken from Ref.~\cite{hegde_environment-dependent_2014}. Note that in contrast to Hegde and coworkers, we do not take off-diagonal intra-atomic matrix elements into account since we have found that such matrix elements can give rise to unphysical results in nanoparticle calculations.

To construct the tight-binding Hamiltonian of a spherical Ag-Au nanoparticle of radius $R$, we start by setting up the Hamiltonian of a bulk alloy, choose an atom as the center of the nanoparticle and then remove all inter- and intraatomic matrix elements involving atoms whose distance from the atom at the center is larger than $R$. Note that in the procedure the onsite energies of atoms at the surface retain a bulk-like character. 

To calculate the matrix element of the total potential, we use that 
\begin{equation}
    \langle{j,\beta}|\hat{\Phi}_{\text{tot}}(\omega)|i,\alpha\rangle = \Bigg(-eE_{0} \frac{3\epsilon_m}{2\epsilon_m+\epsilon(\omega)} z_j \Bigg) \delta_{ij} \delta_{\alpha\beta},
    \label{matrix_element}
\end{equation}
where $\alpha$ and $\beta$ label tight-binding basis functions and $z_j$ denotes the $z$-coordinate of atom $j$. 

Evaluating Eq.~\ref{HCG_eqn} is challenging for large nanoparticles since the calculation of the nanoparticle wavefunctions and energies requires a diagonalization of the tight-binding Hamiltonian. To avoid this, we use the kernel polynomial method, see Refs.~\cite{jin_plasmon-induced_2022,weise_kernel_2006} for details. Additional details about implementation are provided in the Supplementary Information.

The total hot-electron generation rate (which is equal to the total hot-hole generation rate) is given by 
\begin{equation}
    N_{tot}(\omega) = \int_{E_f}^{\infty} N_e(E,\omega)dE,
    \label{total_hot_carriers}
\end{equation}
where $E_f$ denotes the Fermi energy.

To determine the total number of holes (and electrons) $N_{\text{inter}}(\omega)$ generated by interband transitions, i.e. by a transition from a d-band state to an sp-band state, we first determine the d-band onset $E_d$ from the nanoparticle density of states (DOS) and then evaluate
\begin{align}
    N_{\text{inter}}(\omega) =  \int_{-\infty}^{E_d} N_h(E,\omega)dE
    \label{total_inter}.
\end{align}

To calculate the total number of electrons (and holes) $N_\text{intra}(\omega)$ generated from intra-band transitions, i.e. by a transition from an occupied sp-band state to an empty sp-band state, we evaluate
\begin{align}
    N_{\text{intra}}(\omega) = \int_{E_d+\hbar\omega+\delta}^{\infty} N_e(E,\omega)dE
    \label{total_intra}
\end{align}
with $\delta = 0.05$~eV chosen to exclude any contribution from interband transitions.

\begin{acknowledgement}
S.J. and J.L. acknowledge funding from the Royal Society through a Royal Society University Research Fellowship URF/R/191004 and also from the EPSRC programme grant EP/W017075/1. 
\end{acknowledgement}

\begin{suppinfo}
The supplementary information contains computational details and also the tight-binding parameters for Au, Ag and Au-Ag nanoparticles.

\end{suppinfo}

\bibliography{references}

\end{document}